\documentclass[reprint,amsmath,amssymb,aps,showkeys]{revtex4-2}

\usepackage{slashed}
\usepackage{amsmath}

\usepackage{graphicx}
\usepackage{caption}
\usepackage{subcaption}
\usepackage[justification=raggedright,format=hang]{caption}
\usepackage{dcolumn}
\usepackage{bm}
\usepackage{hyperref}
\usepackage[mathlines]{lineno}
\usepackage{xcolor}
\usepackage[makeroom]{cancel}
\usepackage{upgreek}
\usepackage[normalem]{ulem}

\begin{document}


\title{Torsional Carroll Gravity} 

\author{Patrick Concha$^{\ast,\star}$, Nelson Merino$^{\dagger,\ddagger}$, Lucrezia Ravera$^{\bullet,\diamond,\star}$, Evelyn Rodríguez$^{\ast,\star}$}
\email{patrick.concha@ucsc.cl, \\
nemerino@unap.cl, \\ 
lucrezia.ravera@polito.it, \\
erodriguez@ucsc.cl}
\affiliation{
$^{\ast}$Departamento de Matemática y Física Aplicadas, Universidad Católica de la Santísima Concepción, Alonso de Ribera 2850, Concepción, Chile. \\
$^{\star}$Grupo de Investigación en Física Teórica, GIFT, Universidad Católica de la Santísima Concepción, Alonso de Ribera 2850, Concepción, Chile. \\
$^{\dagger}$Instituto de Ciencias Exactas y Naturales, Universidad Arturo Prat, Avenida Playa Brava 3256, 1111346, Iquique, Chile. \\
$^{\ddagger}$Facultad de Ciencias, Universidad Arturo Prat, Avenida Arturo Prat Chacón 2120, 1110939, Iquique, Chile. \\
$^{\bullet}$DISAT, Politecnico di Torino, Corso Duca degli Abruzzi 24, 10129 Torino, Italy. \\
$^{\diamond}$Istituto Nazionale di Fisica Nucleare (INFN), Section of Torino,
Via P. Giuria 1, 10125 Torino, Italy. \\
}

\date{\today}

\begin{abstract}
The ultra-relativistic (Carrollian) regime of gravity has recently emerged as a fertile framework for exploring holography, non-Lorentzian symmetries, and geometric limit of General Relativity. In this letter, we establish the presence of a non-vanishing torsion within three-dimensional Carrollian gravity by constructing the Carrollian Mielke-Baekler (C-MB) gravity theory in its Chern-Simons formulation, obtained as the ultra-relativistic limit of the relativistic Mielke-Baekler model. The resulting C-MB theory features non-zero temporal torsion and curvature, together with spatial curvature, providing the most general three-dimensional Carrollian gravity model with these properties. 
Temporal torsion affects non-affinity of null generators and boundary dynamics.
Several known ultra-relativistic gravity theories arise as particular limits of this framework, highlighting its unifying character.
\end{abstract}

\maketitle

\section{Introduction}\label{Introduction}

Carrollian symmetries have emerged across a broad range of physical settings, from tensionless strings and warped CFTs \cite{Hofman:2014loa,Bagchi:2013bga,Bagchi:2018wsn} to asymptotic symmetries, flat holography, and black-hole horizons \cite{Duval:2014uva,Hartong:2015xda,Hartong:2015usd,Bagchi:2016bcd,Donnay:2019jiz,Ciambelli:2019lap,Grumiller:2019fmp,Perez:2021abf,Donnay:2022aba,Perez:2022jpr,Fuentealba:2022gdx,Saha:2023hsl,Nguyen:2023vfz,deBoer:2023fnj,Ecker:2023uwm,Donnay:2023mrd,Bagchi:2023cen}. The ultra-relativistic limit of General Relativity has therefore become a powerful tool for uncovering non-Lorentzian geometric structures, positioning Carrollian gravity as a promising bridge between high-energy theory, holography, and non-Lorentzian geometry.

At the level of gravity, the Carrollian theory arises from sending the speed of light to zero in the Einstein-Hilbert action \cite{Bergshoeff:2017btm}. In three spacetime dimensions, it can alternatively be formulated as a Chern-Simons (CS) theory based on the Carroll algebra \cite{Matulich:2019cdo,Ravera:2019ize,Ali:2019jjp,Concha:2023bly}, the ultra-relativistic contraction of the Poincaré algebra \cite{Levy-Leblond:1965dsc}. The three-dimensional CS framework offers a powerful arena for probing structural aspects of gravity, including black-hole physics and thermodynamics  \cite{Banados:1992wn}. Matter couplings to Carroll gravity have also been successfully constructed \cite{Hansen:2021fxi,Baiguera:2022lsw,Bergshoeff:2024ilz} performing systematic Carrollian expansions of relativistic models.

By contrast, the role of torsion in the ultra-relativistic regime remains poorly understood. While torsion is well established in Newton-Cartan gravity, arising naturally from gauging the Schrödinger algebra \cite{Bergshoeff:2014uea} and appearing in contexts such as Lifshitz holography \cite{Christensen:2013lma} and the Quantum Hall Effect \cite{Christensen:2013rfa}, no fully general formulation of torsional Carroll gravity exists. Current Carrollian models either enforce vanishing torsion or impose curvature constraints, preventing a complete geometric description. In particular, no three-dimensional Carrollian gravity theory \emph{simultaneously} accommodating non-trivial torsion and curvature has been constructed to date, leaving torsional effects in the ultra-relativistic regime essentially unexplored.

In this letter, we fill this gap by constructing the first three-dimensional Carrollian gravity theory that simultaneously exhibits non-vanishing torsion and curvature. Starting from the relativistic Mielke-Baekler (MB) model \cite{Mielke:1991nn, Baekler:1992ab}, a topological gravity theory that naturally incorporates torsion, we perform its ultra-relativistic contraction to obtain a Carrollian Mielke-Baekler (C-MB) theory in the CS formulation. The resulting C-MB model features non-zero temporal Carroll torsion, temporal curvature, and spatial curvature, thereby providing the most general and structurally complete Carrollian gravity theory to date. We further show that diverse known ultra-relativistic gravity models arise as specific limits of our construction, revealing its unifying character, and discuss some relevant physical implications of our C-MB theory on null hypersurfaces and boundary dynamics. 
This work places torsional Carroll gravity on solid geometric foundations and opens a new arena for exploring torsional effects in non-Lorentzian geometry and ultra-relativistic gravitational dynamics.

\section{Mielke-Baekler gravity}\label{MBgrav}

The most general three-dimensional gravitational Lagrangian constructed from the vielbein $E^{A}$ and the spin connection $W^{A}$, originally introduced by Mielke and Baekler \cite{Mielke:1991nn, Baekler:1992ab}, reads
\begin{align}
    L_{\text{MB}}[E^{A},W^{A}]&=\sigma_{0}L_{0}[E^{A}]+\sigma_{1} L_{1}[E^{A},W^{A}] \notag\\
    &+\sigma_{2} L_{2}[W^{A}]+\sigma_{3} L_{3}[E^{A},W^{A}]\,,\label{origMBLagr}
\end{align}
where the $\sigma_i$'s, $i=0,1,2,3$, are independent constants and
\begin{align}
    L_0[E^A] & := \frac{1}{3} \epsilon_{ABC} E^A E^B E^C \,,\notag \\
    L_1 [E^A,W^A] & := 2 E_A R^A \,, \notag\\
    L_2 [W^A] & := W^A dW_A + \frac{1}{3} \epsilon^{ABC} W_A W_B W_C \,, \notag\\
    L_3 [E^A,W^A] & := E_A T^A \label{LagMB}\,.
\end{align}
Here $A=0,1,2$ are Lorentz indices and $\epsilon_{ABC}$ is the Levi Civita tensor satisfying $\epsilon_{012}=-\epsilon^{012}=1$. The Lorentz curvature and torsion 2-forms are
\begin{align}
    R^A & = dW^A + \frac{1}{2} \epsilon^{ABC} W_B W_C \,,\notag \\
    T^A & = dE^A + \epsilon^{ABC} W_B E_C \,.
\end{align}
Note that $L_0$ provides a cosmological constant term, $L_1$ corresponds to the Einstein-Hilbert contribution, $L_2$ is the CS gravitational (or exotic) term \cite{Witten:1988hc}, and $L_3$ is a ``translational" CS structure. 
MB gravity has been the subject of extensive study \cite{Blagojevic:2003uc,Blagojevic:2003vn, Blagojevic:2004hj,Klemm:2007yu,Cvetkovic:2007sr,Peleteiro:2020ubv,Concha:2023ejs}, ranging from its  black-hole solutions to its asymptotic symmetries, holographic behavior, supersymmetric extensions, and interactions with higher-spin gauge fields.

The MB Lagrangian admits a CS formulation based on the so-called \emph{MB algebra}, spanned by the generators of Lorentz transformations $J_{A}$ and translations $P_{A}$. Its non-zero commutation relations are given by \cite{Geiller:2020edh}
\begin{align}
        & [J_A, J_B] = \epsilon_{ABC} J^C \,,\notag \\
        & [J_A, P_B] = \epsilon_{ABC} P^C \,,\notag \\
        & [P_A, P_B] = \epsilon_{ABC} \left( p J^C + q P^C \right) \,. \label{MBalgebra} 
\end{align}
Here, we adopt the Minkowski metric $\eta_{AB} = (-1, 1, 1)$, and the dimensional parameters $p$ and $q$ determine the cosmological constant,
\begin{align}
\label{cosmconstMB}
    \Lambda := - \left(p + \frac{q^2}{4} \right)\,.
\end{align}

The MB algebra admits a non-degenerate invariant bilinear form given by
\begin{align}
    & \langle J_A J_B \rangle = \sigma_2 \eta_{AB} \,, \notag \\
    & \langle J_A P_B \rangle = \sigma_1 \eta_{AB} \,, \notag \\
    & \langle P_A P_B \rangle = \left( p \sigma_2 + q \sigma_1 \right) \eta_{AB} \,, \label{MBinvtens}
\end{align}
with the non-degeneracy condition
\begin{equation}
\sigma_1\neq \sigma_2\left(\frac{q}{2}\pm\sqrt{p+\frac{q^2}{4}}\right)\,. \label{NDeg}
\end{equation}
To obtain the CS formulation of the MB model \cite{Blagojevic:2003vn,Cacciatori:2005wz,Giacomini:2006dr,Geiller:2020edh} (see also \cite{Concha:2023ejs}), 
we start by introducing the gauge connection one-form $A$ for the MB algebra,
\begin{align}
\label{AMB}
    A=W^A J_A + E^A P_A \,,
\end{align}
with the associated curvature two-form,
\begin{align}
\label{FMB}
    F= \mathcal{R}^A (W) J_A + \mathcal{R}^A (E) P_A \,,
\end{align}
where
\begin{equation}
\label{RWandRE-MB}
\begin{aligned}
    \mathcal{R}^A(W) & := dW^A + \frac{1}{2} \epsilon^{ABC} W_B W_C + \frac{p}{2} \epsilon^{ABC} E_B E_C \,, \\
    \mathcal{R}^A(E) & := d E^A + \epsilon^{ABC} W_B E_C + \frac{q}{2} \epsilon^{ABC} E_B E_C \,. 
\end{aligned}
\end{equation}
By inserting the gauge connection one-form \eqref{AMB} and the invariant bilinear form, whose non-vanishing components are given by \eqref{MBinvtens}, into the standard CS Lagrangian expression,
\begin{align}
    L_{\text{CS}}=\Big\langle AdS+\frac{2}{3}A^{3}\Big\rangle\,,
\end{align}
we obtain the CS Lagrangian for the MB algebra \eqref{MBalgebra}:
\begin{align}
    L_{\text{CS}}[A] & :=  (p \sigma_1 + q \sigma_3) L_0 [E^A] + \sigma_1 L_1 [E^A,W^A] \notag \\
    & + \sigma_2 L_2 [W^A] + \sigma_3 L_3 [E^A,W^A] \,,\label{MBCSLagr}
\end{align}
where the constants satisfy
\begin{align}
    \label{rel_s3-s2s1}
    \sigma_3 = p \sigma_2 + q \sigma_1\,,
\end{align}
and the Lagrangian terms are defined in \eqref{LagMB}. In particular, the CS Lagrangian \eqref{MBCSLagr} reproduces the original MB expression \eqref{origMBLagr} upon identifying
\begin{align}
    \sigma_0 = p \sigma_1 + q \sigma_3 \,.
\end{align}
One then finds, up to boundary contributions \cite{Geiller:2020edh},
\begin{align}
    L_{\text{CS}}[A] = L_{\text{MB}} [E^A,W^A] \,.
\end{align}
Let us remark that in \eqref{MBCSLagr} the parameters $\sigma_i$, $i=0,1,2,3$, can be fixed allowing one to recover Einstein-Hilbert gravity, teleparallel gravity, and the ``exotic" Witten gravity model \cite{Witten:1988hc}.

Requiring the invariant tensor to be non-degenerate, i.e., for $\sigma_1$ and $\sigma_2$ satisfying \eqref{NDeg}, the field equations derived from the CS Lagrangian \eqref{MBCSLagr} reduce to the vanishing of the curvature components \eqref{FMB}, $\mathcal{R}^A(W)=0$ and $\mathcal{R}^A(E)=0$. These are equivalently expressed as
\begin{align}
\label{relMBeq}
    & 2 R^A + p \epsilon^{ABC} E_B E_C = 0 \,, \notag\\
    & 2 T^A + q \epsilon^{ABC} E_B E_C = 0 \,, 
\end{align}
which coincides with the field equations of the MB Lagrangian \eqref{origMBLagr}, assuming $\sigma_1^2-\sigma_2 \sigma_3 \neq 0$.

\section{Carrollian Mielke-Baekler Gravity}\label{CarMBgrav}

We now construct the Carrollian counterpart of the MB CS gravity theory. Starting from the MB algebra \eqref{MBalgebra}, we decompose the generators $\lbrace{J_0, J_a, P_0,P_a \rbrace}$, with the Lorentz index split as $A=(0,a)$ and $a=1,2$. The Carrollian limit is obtained through the rescaling of the generators as
\begin{align}
    J_{0}&\rightarrow J \,, & P_{a}&\rightarrow P_{a}\,,\notag\\
    P_0 &\rightarrow \sigma H \,, & J_a &\rightarrow \sigma K_a \,, &  q\rightarrow 1/\sigma \, q \,,
    \label{resc}
\end{align}
followed by the limit $\sigma \rightarrow \infty$. The $\sigma$ parameter is related to the speed of light $c$ as $\sigma\rightarrow 1/c$, so that the limit $\sigma\rightarrow\infty$ corresponds to the
ultra-relativistic limit $c\rightarrow 0$. The resulting non-vanishing commutation relations of the contracted algebra define the \emph{Carroll Mielke-Baekler algebra} ($\mathfrak{car}_{\texttt{MB}}$ algebra in the following) and satisfy:
\begin{align}
    [J,K_a] &= \epsilon_{ab} K_b \,, \notag \\
    [K_a,P_b] &= - \epsilon_{ab} H \,, \notag \\
    [J,P_a] &= \epsilon_{ab} P_b \,, \notag \\
    [P_a,P_b] &= - \epsilon_{ab} \left(p J + q H \right) \,,\notag  \\
    [H,P_a] &= p\, \epsilon_{ab} K_b \,. \label{cmbalgebra}
\end{align}
Here, the generators $\lbrace{ J,K_a,H,P_a \rbrace}$ are, respectively, spatial rotations, Carrollian boosts, time translations, and space translations. 
The first three relations in \eqref{cmbalgebra} reproduce the Carroll algebra \cite{Levy-Leblond:1965dsc,Bacry:1968zf}, obtained as an ultra-relativistic contraction of the Poincaré algebra. It also corresponds to the flat limit ($\ell \rightarrow \infty$)  of the AdS-Carroll algebra \cite{Bergshoeff:2015wma,Matulich:2019cdo}. The latter is recovered from \eqref{cmbalgebra} by setting $p=1/\ell^2$ and $q=0$. Non-zero values of $p$ and $q$ introduce  non-vanishing temporal components of both Carrollian torsion and Carrollian curvature. 

Using the decomposition $A=(0,a)$ in the non-vanishing components of the invariant bilinear form \eqref{MBinvtens} and applying the rescaling \eqref{resc} together with
\begin{align}
    \sigma_1 &\rightarrow \sigma \tilde{\sigma}_1 \,, & \sigma_2 &\rightarrow \tilde{\sigma}_2 \,, \label{sigmasresc}
\end{align}
we obtain, after taking the limit $\sigma \rightarrow \infty$, the non-vanishing components of the $\mathfrak{car}_{\texttt{MB}}$ invariant tensor:
\begin{align}
    \langle J J \rangle &= - \tilde{\sigma}_2 \,,  \notag \\
    \langle K_a P_b \rangle &= \tilde{\sigma}_1 \delta_{ab} \,, \notag \\
    \langle J H \rangle &= - \tilde{\sigma}_1 \,, \notag \\
    \langle P_a P_b \rangle &= \left( p \tilde{\sigma}_2 + q \tilde{\sigma}_1 \right) \delta_{ab} \,. \label{CMBIT}
\end{align}
Since $q \rightarrow 1/\sigma \, q$, one finds that $\sigma_3 = p \sigma_2 + q \sigma_1 \rightarrow \tilde{\sigma}_3 = p \tilde{\sigma}_2 + q \tilde{\sigma}_1 $, while $\sigma_0 = p \sigma_1 + q \sigma_3 \rightarrow \sigma p \tilde{\sigma}_1 + \mathcal{O}(\sigma^{-1})$. The invariant bilinear form \eqref{CMBIT} is non-degenerate (for $\tilde{\sigma}_1 \neq 0$), meaning that, in this case, the ultra-relativistic limit preserves non-degeneracy without the need to introduce additional auxiliary generators (and corresponding gauge fields). 
This property is crucial for obtaining a well-defined Carrollian CS formulation, ensuring the presence of a kinetic term for each gauge field and field equations given by the vanishing of the corresponding curvature two-forms.

To construct the CS Lagrangian based on the $\mathfrak{car}_{\texttt{MB}}$ algebra, we start by introducing the gauge connection one-form 
\begin{align}
    A &= \tau H + e^a P_a + \omega J + \omega^a K_a \,, \label{ACMB}
\end{align}
and the corresponding curvature two-form $F=dA+\tfrac{1}{2}[A,A]$, which decomposes as
\begin{align}
    F &= R \left(\tau \right) H + R
    ^a \left(e^b \right) P_a + R \left( \omega \right) J + R
    ^a \left(\omega^b \right) K_a \,.\label{FCMB}
\end{align}
The explicit expressions for the components of the curvature are
\begin{align}
    R \left( \omega \right) & :=  d \omega + \frac{p}{2} \epsilon^{ac} e_a e_c \,, \notag \\
    R^a \left(\omega^b \right) & := d \omega^a +  \epsilon^{ac}  \omega \omega_c + p \epsilon^{ac} \tau e_c \,,\notag \\
    R \left(\tau \right) & := d \tau  + \epsilon^{ac} e_a \omega_c + \frac{q}{2} \epsilon^{ac} e_a e_c  \,, \notag\\
    R^a \left(e^b \right) & := d e^a + \epsilon^{ac} \omega e_c \,. \label{curvCMB}
\end{align}
These curvature two-forms are obtained as the Carrollian limit of the relativistic MB curvature 2-forms \ref{RWandRE-MB}, after implementing the index splitting $A=(0,a)$. We observe that setting $p=q=0$ one recovers the standard Carroll curvature two-forms.

Using the non-vanishing components of the invariant tensor for the $\mathfrak{car}_{\texttt{MB}}$ algebra \eqref{CMBIT} together with the connection one-form \eqref{ACMB}, the ultra-relativistic CS Lagrangian based on the $\mathfrak{car}_{\texttt{MB}}$ algebra \eqref{cmbalgebra} takes the form
    \begin{align}
        L_{\text{C-MB}} &=  - \, \tilde{\sigma}_0 \epsilon^{ac} \tau e_a e_c + 2 \tilde{\sigma}_1 \left[ e_a \tilde{R}^a\left(\omega^b \right) - \tau \tilde{R}\left(\omega\right) \right] \notag\\
        & - \tilde{\sigma}_2 \omega \tilde{R}\left(\omega\right) + \tilde{\sigma}_3 e_a R^a \left( e^b \right) \,,\label{C-MBCSLagr}
    \end{align}
where we have defined
    \begin{align}
    \tilde{R}\left(\omega\right) &:= d \omega \,, \notag\\
    \tilde{R}^a\left(\omega^b \right) &:= d \omega^a + \epsilon^{ac} \omega \omega_c \,,
    \end{align}
and identified $\tilde{\sigma}_0=p\tilde{\sigma}_{1}$ and $\tilde{\sigma}_3=p\tilde{\sigma}_2+q\tilde{\sigma}_1$. The resulting model, which we denote as the C-MB gravity, constitutes the ultra-relativistic counterpart of the MB gravity Lagrangian \eqref{origMBLagr}. In particular, the Lagrangian \eqref{C-MBCSLagr} can be interpreted as the ultra-relativistic analogue of the cosmological constant term, the Einstein-Hilbert contribution, the exotic term \cite{Witten:1988hc}, and the ``translational" CS structure. Remarkably, specific values of the parameters $p$ and $q$ reproduce distinct three-dimensional Carrollian CS gravity models, as summarized in Table \ref{tab:C-MBpq}.

\begin{table}[ht]
\caption{Carrollian gravity theories within the C–MB framework.}
\label{tab:C-MBpq}
\begin{ruledtabular}
\begin{tabular}{lccc}
Carrollian-type gravity& $p$ & $q$ \\
\hline
Ultra-relativistic torsional gravity & 0 & $-2/\ell$ \\
AdS-Carroll gravity & $1/\ell^2$ & 0 \\
Carroll gravity & 0 & 0 \\
\end{tabular}
\end{ruledtabular}
\end{table}

Let us remark that the C-MB Lagrangian \eqref{C-MBCSLagr} can be obtained directly from the relativistic MB Lagrangian  \eqref{origMBLagr} by performing an appropriate ultra-relativistic contraction. This follows from the rescaling of the relativistic MB coupling constants,
\begin{align}
    \sigma_{0}&\rightarrow \sigma \tilde{\sigma}_{0}\,, & \sigma_{1}&\rightarrow \sigma\tilde{\sigma}_{1}\,, \notag\\
    \sigma_{2}&\rightarrow \tilde{\sigma}_{2} \,, & \sigma_{3}&\rightarrow \tilde{\sigma}_{3}\,,
\end{align}
together with the following rescaling of the relativistic gauge fields:
\begin{align}
    W^{0}&\rightarrow \omega\,, &  E^{a}&\rightarrow e^{a}\,, \notag\\
    E^{0}&\rightarrow \sigma^{-1}\tau\,, & W^{a}&\rightarrow \sigma^{-1} \omega^{a}\,, 
\end{align}
followed by the limit $\sigma\rightarrow\infty$. While this contraction reproduces the C-MB Lagrangian \eqref{C-MBCSLagr} in a systematic manner, it does not reveal the underlying symmetry algebra of the resulting ultra-relativistic gravity theory.

The C-MB gravity model describes the most general three-dimensional Carrollian gravity theory featuring non-vanishing Carrollian curvature and non-zero temporal components of Carrollian torsion. For arbitrary values of $p$ and $q$ (and assuming the non-degeneracy condition $\tilde{\sigma}_1\neq 0$), the field equations of the theory are given by the vanishing of the C-MB curvature two-forms \eqref{curvCMB}:
\begin{equation}
\begin{aligned}
    \delta \tau : & \quad R \left(\omega \right) = 0 \,, \\
    \delta \omega : & \quad R \left(\tau \right) = 0 \,, \quad \text{as } R \left(\omega \right) =0\,, \\
    \delta \omega_a : & \quad R^a \left( e^b \right) = 0 \,, \\ 
    \delta e_a : & \quad R^a \left( \omega^b \right) = 0 \,, \quad \text{as } R^a \left(e^b \right) =0\,.
\end{aligned}
\end{equation}
A key feature is the vanishing of $R\left(\tau\right)$, which implies a non-zero ``temporal component" of the Carrollian torsion:
\begin{align}
\label{temporaltorscomp}
    d\tau+\epsilon^{ac}e_{a}\omega_{c} &\neq 0\,.
\end{align}
In the ultra-relativistic teleparallel version ($p=0$ and $q=-2/\ell$), the cosmological constant acts as a source for this temporal Carrollian torsion. Similarly, for $p\neq0$, $R\left(\omega\right)=0$ and $R^a \left( \omega^b \right)=0$ imply the presence of non-vanishing spatial curvature in the vacuum of the theory, whereas the spatial torsion $R^a \left( e^b \right)$ vanishes on-shell. Each of these sub-cases has a physical interest on its own and defines a distinct Carrollian gravitational sector.

\subsection{Physical implications of torsional Carroll gravity}\label{Physical implications of torsional Carroll gravity}

Carrollian gravity theories are known to play a central role in the description of null hypersurfaces and of boundary geometries in UR and asymptotically flat spacetimes. 
The C-MB model constructed in this Letter presents 
an intrinsic temporal torsion.
We now highlight the physical implications of this geometric structure, focusing on its impact on the geometry of Carrollian horizons and on the structure of boundary dynamics, relevant for, e.g., flat holography. 
In particular, we show that temporal Carrollian torsion acquires a direct physical interpretation in terms of horizon non-affinity and defines a new geometric datum at null infinity.

\subsubsection{Effects of Carrollian torsion on null hypersurfaces}

Carrollian geometry naturally emerges as the intrinsic geometry of null hypersurfaces, such as black hole event horizons or null infinity \cite{Donnay:2019jiz,Hopfmuller:2016scf,Ciambelli:2019lap,Blitz:2023yam}. 
In this framework, the Carrollian clock 1-form $\tau$ is associated with the null generators of the hypersurface, while the spatial vielbein $e^a$ encodes the geometry of its spatial sections.

In the presence of temporal Carrollian torsion, this structure naturally endows null hypersurfaces with an additional \emph{geometric charge} contribution (in our C-MB theory, the torsional boundary datum/source $q$), fixed by the underlying Carrollian symmetry data.

The temporal Carrollian torsion
$T_{\mathrm{T}} := d\tau + \epsilon^{ab} e_a \omega_b$ measures the failure of the null generators to be affinely parametrized. 
This is the Carrollian analogue of the standard relation
\begin{align}
    k^\nu \nabla_\nu k^\mu = \kappa k^\mu
\end{align}
for null generators $k^\mu$,
with $T_\text{T}$ here playing the role of a geometric realization of the non-affinity $\kappa$.
The torsion deformations, and their UR effects, are encoded in the covariant derivative acting on null generators.
In the C-MB theory, the field equations impose
\begin{align}
\label{torsconstr}
    T_{\mathrm{T}} = -\frac{q}{2} \epsilon^{ab} e_a e_b ,
\end{align}
so that the non-affinity of the null generators is fixed by the torsional coupling $q$.

This provides a geometric and symmetry-based Carrollian interpretation of the surface gravity of non-extremal horizons, independent of the existence of a specific global black hole solution. 
In this sense, the parameter $q$ controls torsional contributions associated with horizon acceleration, while the parameter $p$ governs the intrinsic curvature of the spatial horizon sections. 
The C-MB theory therefore endows null hypersurfaces with a torsional charge, analogous, at the level of boundary data, to a chemical potential for horizon acceleration.

\subsubsection{Implications on boundary dynamics}

Carrollian geometries provide the natural boundary data for asymptotically flat spacetimes, where the dual dynamics is governed by Carrollian field theories with BMS-type symmetries. 
Within the C-MB framework, intrinsic temporal Carrollian torsion enlarges the space of admissible boundary geometries beyond the torsion-free case, while remaining compatible with a CS variational principle.

From the CS formulation, the on-shell variation of the action reduces to a boundary contribution,
\begin{align}
\delta I_{\text{C-MB}}\big|_{\text{on-shell}}
= \int_{\partial\mathcal{M}} \langle \delta A \wedge A \rangle ,
\end{align}
where $\mathcal{M}$ is the three-dimensional bulk manifold and $\partial\mathcal{M}$ its boundary. Consistent boundary conditions and the associated boundary symplectic structure are therefore entirely determined by the invariant bilinear form \eqref{CMBIT}.

From the latter, the boundary symplectic structure induced by the CS action identifies the canonical pairs $(\omega,\tau)$ and $(\omega^{a},e_{a})$, with symplectic weights proportional to $\tilde{\sigma}_{1}$. 
In particular, the non-vanishing pairing $\langle JH\rangle=-\tilde{\sigma}_{1}$ implies that the Carrollian clock 1-form $\tau$ is canonically conjugate to spatial rotations at the boundary, while the pairing $\langle K_{a}P_{b}\rangle=\tilde{\sigma}_{1}\delta_{ab}$ governs the boost-translation sector.

The variation of the surface charge associated with a Lie-algebra-valued gauge parameter $\lambda$ generating infinitesimal gauge transformations of the CS connection takes the standard CS form
\begin{align}
\delta Q[\lambda] = \int_{\partial\mathcal{M}} \langle \lambda\, \delta A \rangle .
\end{align}
In the C-MB theory, the temporal Carrollian torsion constraint \eqref{torsconstr}
enters the boundary variational principle as a fixed background source. 
As a consequence, the surface charge associated with Carrollian time translations acquires an explicit $q$-dependent contribution, persisting even in the absence of bulk curvature.

For generic values of $q$, the residual gauge transformations preserving the torsional boundary condition therefore form a deformation of the standard Carrollian (or $\mathfrak{bms}_3$-type) algebra.
This deformation persists in the flat limit $p=0$, indicating that intrinsic temporal Carrollian torsion defines a genuinely new sector of Carrollian boundary dynamics. 
The parameter $p$, which controls spatial curvature, further allows for curved Carrollian boundary geometries, providing a unified framework encompassing flat, AdS-Carroll, and torsional Carrollian holography.

\section{Discussion}\label{Conclusion}

We have constructed the Carrollian Mielke–Baekler (C-MB) gravity theory as the ultra-relativistic limit of the relativistic Mielke–Baekler model, obtaining the most general three-dimensional Carrollian gravity theory featuring non-vanishing temporal torsion together with temporal and spatial curvature. This result provides the first consistent framework in which torsional effects can be systematically explored in the ultra-relativistic regime and unifies several previously known Carrollian theories as particular subcases of a single geometric model.

The CS formulation of the C-MB model  offers a versatile starting point for future investigations. In particular, it provides a natural setting for studying asymptotic symmetries, conserved charges, and possible torsion-induced deformations of Carrollian or $\mathfrak{bms}_3$-type algebras. Our construction also opens the door for exploring the holographic correspondence in ultra-relativistic regimes, including potential extensions of the conformal field theory  side. Extensions to supersymmetry, matter couplings, and higher-dimensional generalizations constitute further promising directions. Together, these avenues may shed new light on the role of torsional non-Lorentzian geometries in geometric limits of General Relativity and quantum gravity.

The C-MB theory modifies the topological description of null hypersurfaces in the UR regime, by extending Carrollian horizon models to include both curvature and intrinsic temporal torsion. 
Here, torsion affects non-affinity of null generators and acts as a geometric source for horizon acceleration. 
This framework naturally enables the study of torsion-dependent modifications of horizon symmetries, conserved charges, and near-horizon dynamics. 
As an illustration, one may consider the UR limit of the MB BTZ solution. 
Detailed derivations of Carrollian black holes with torsion, boundary conditions, surface charges, and the resulting deformed Carrollian symmetry algebra are left for future work.

\begin{acknowledgments}
This work was funded by the National Agency for Research and Development
ANID - FONDECYT grants 1250642 and 1231133. This work was
supported by the Research project Code DIREG 04/2025 (P.C.) of the Universidad Católica de la Santísima
Concepción (UCSC), Chile. P.C., L.R and E.R. would also like to thank
the Dirección de Investigación and Vice-rectoría de Investigación
of the Universidad Católica de la Santísima Concepción, Chile, for
their constant support. L.R. is supported by the research grant PNRR Young Researchers, funded by MIUR, MSCA Seal of Excellence (SoE), CUP E13C24003600006, ID SOE2024$\_$0000103, project GrIFOS, of which this paper is part.
\end{acknowledgments}

\bibliography{MBCarroll.bib}

\end{document}